\documentclass[10pt,pre,a4paper,twocolumn,superscriptaddress,showpacs]{revtex4-1}
\usepackage{amsmath}
\usepackage{amssymb}
\usepackage{amsthm}
\usepackage{mathtools}
\usepackage{graphicx}
\usepackage{grffile}
\usepackage{color}
\usepackage[dvipsnames]{xcolor}

\newcommand{\xx}[1]{{ #1}}

\renewcommand{\vec}[1]{\mathbf{#1}}

\begin{document}

\author{Claus Heussinger} \affiliation{Institute for theoretical
  physics, Georg August University G\"ottingen, Friedrich Hund Platz 1,
  37077 G\"ottingen, Germany} \title{Packings of frictionless
  spherocylinders}

\begin{abstract}
  We present simulation results on the properties of packings of
  frictionless spherocylindrical particles. Starting from a random
  distribution of particles in space, a packing is produced by
  minimizing the potential energy of inter-particle contacts until a
  force-equilibrated state is reached. For different particle aspect
  ratios $\alpha=10\ldots 40$, we calculate contacts $z$, pressure as
  well as bulk and shear modulus.  Most important is the fraction
  $f_0$ of spherocylinders with contacts at both ends, as it governs the jamming
  threshold $z_c(\alpha)=8+2f_0(\alpha)$. These results highlight
  the important role of the axial ``sliding'' degree of freedom of a
  spherocylinder, which is a zero-energy mode but only if no
  end-contacts are present.
\end{abstract}

\maketitle

\section{Introduction}

Packing and flow of granular particles has been the subject of intense
research over many years. Even though granular particles generally
have irregular, non-spherical shapes, work has mainly been concerned
with the properties of systems of spherical particles. A few studies
deal with flow properties of non-spherical particles of different
shapes, e.g. cube-like shapes~\cite{C5SM00729A,C6SM00205F}, needles
\cite{tapia_shaikh_butler_pouliquen_guazzelli_2017,egres05,PhysRevE.84.031408}
or simulations with polygons~\cite{PhysRevE.66.021301} or
ellipsoids~\cite{campbell_ellipse,trulsson_2018}. Coupling of shear
flow to rotational degrees of freedom leads to non-trivial alignment
properties~\cite{PhysRevLett.108.228302,PhysRevE.97.012905,nath19:_rheol},
even in the limit of nearly spherical
particles~\cite{PhysRevE.100.032906,somfai}.

The jamming properties of nearly spherical, ellipsoidal or
spherocylindrical particles have been discussed in some detail.
Slightly non-spherical particles can use space more efficiently and
pack at optimized, higher
densities~\cite{willi03,PhysRevE.75.051304}. Frictional interaction
forces, however, seem to act against this packing
optimization~\cite{nath19:_rheol}. Jamming of nearly-spherical
particles is complicated by the special role of the rotational degrees
of freedom
~\cite{zeravcic2009EPL,PhysRevLett.102.255501,PhysRevE.97.012909,hecke}.  In
the sphere limit rotational degrees of freedom are lost. However,
already for nearly-spherical ellipsoids the jamming transition is
modified as rotational and translational degrees of freedom form
separate bands that only weakly mix~\cite{zeravcic2009EPL}.

Here we are concerned with packings of frictionless spherocylindrical
particles. We are primarily interested in long, high-aspect-ratio
($\alpha$) particles and the question how the limiting behavior to
$\alpha\to\infty$ looks like. Infinitely long spherocylinders have a
symmetry related to translations along their axis (``sliding''), which
cannot be constrained by any interactions. The question is in how far
this symmetry is still visible in the jamming of
nearly-infinite-length, i.e. finite-length spherocylinders, just as
rotational symmetry is still visible in nearly-spherical particles.

Previous work on high-aspect ratio particles has dealt with the
jamming of elastic fibers~\cite{rod05,PhysRevLett.118.068002}, or
rods~\cite{willi03,Pournin2005,w.tavares03:_influen_monte_carlo,10.1007/978-3-540-69387-1_14}. Frictional
interactions have been seen to strongly affect the response to shear
deformations~\cite{PhysRevE.80.016115}, possibly leading to a
gravity-induced yielding transition as a function of particle
length~\cite{PhysRevE.82.011308,PhysRevE.85.061304}.

We will discuss the jamming properties of static packings of soft
spherocylinders, as well as their response to small bulk and shear
deformations. We will see that the crossover to infinite length is
governed by the fraction of particles which have their axial
translation constrained by contacts at the spherocylinder ends. This
fraction vanishes when $\alpha\to\infty$.

\section{Model}

We study three-dimensional (3d) packings of spherocylindrical
particles. Each spherocylinder (SC) $i=1\ldots N$ consists of a
cylindrical part of length $\ell_i$ and two hemispherical caps of
diameter $d_i$ at the two ends. The center-line of the cylinder is
called the backbone. The volume of a SC is thus
\begin{eqnarray}\label{eq:volume_sc}
  V_{\rm sc} &=& (\pi/6)d^3+(\pi/4)d^2\ell\,.
\end{eqnarray}
The particles interact via repulsive contact forces similar to those
from models for spheres. A contact between particles $i$ and $j$ is
established whenever the shortest distance between the backbones,
$r_{ij} = |\mathbf{r}_{ij}|$, is less than their average diameter
$d_{ij}=(d_{i}+d_{j})/2$. The distance vector can be written as
\begin{equation}\label{eq:rij}
  \mathbf{r}_{ij} = \mathbf{R}_i+\mathbf{\hat n}_is_i -
(\mathbf{R}_j+\mathbf{\hat n}_js_j)\,,
\end{equation}
where $\mathbf{R}_i$ is the position of the center of mass of particle
$i$, $\mathbf{\hat n}_i$ represents the direction of the particle
backbone, and $s_i\in[-\ell_i/2,\ell_i/2]$ is the arclength parameter
along the backbone that specifies where the shortest distance between
$i$ and $j$ is reached. By definition the vector $\mathbf{r}_{ij}$ is
perpendicular to both backbones, except for cases where the shortest
distance is reached at an end of one or both of the SC (i.e.
$s_i=\pm \ell_i/2$). The actual force is applied halfway along the
vector $\mathbf{r}_{ij}$, at the position
$\mathbf{y}_{ij} = \mathbf{\hat n}_is_i+\mathbf{r}_{ij}/2$ away from
the center of mass (with a small correction for unequal-sized
particles). This is, in general, very close to the surface of the two
particles. \xx{The procedure is similar to Ref.~\cite{Pournin2005}.}

The force $\mathbf{f}_{ij}$ on particle $i$ from the contact with $j$
is directed normally to the particle surface. It is calculated as in
the Cundall-Strack model~\cite{cundall79}
\begin{align}
  \mathbf{f}_{ij}&=[-k_n \delta_{ij}- c_nv^n_{ij}]\mathbf{\hat n}_{ij} ,\\
\end{align}
Here, the normal direction
$\mathbf{\hat n}_{ij}=\mathbf{r}_{ij}/r_{ij}$ points from particle $j$
to $i$ at the point of application of the force. The overlap
$\delta_{ij}= d_{ij}-r_{ij}$ is a positive quantity.

The velocity ${v}^n_{ij}$ represents the projection of the relative
velocity $\mathbf{v}_{ij}^{\rm con}$ at the contact. The latter
derives from the center of mass translational $\mathbf{v}_i$ and
rotational motion $\boldsymbol\omega_i$ as
$\mathbf{v}_{ij}^{\rm con} = \mathbf{v}_{i} - \mathbf{v}_{j} -
\mathbf{y}_{ij}\times\boldsymbol\omega_i+\mathbf{y}_{ji}\times\boldsymbol\omega_j$. The
parameter $k_n$ is a spring constant, $c_n$ a viscous damping
constant.

The equations of motion for particle $i$ are
\begin{equation}
 m \mathbf{\ddot r}_i=\sum_j \mathbf{f}_{ij}
\end{equation}
\begin{equation}
 \mathbf{I}_i\cdot \boldsymbol{\dot{\omega}}_i=\sum_j \mathbf{y}_{ij} \times \mathbf{f}_{ij}
\end{equation}
where $\mathbf{I}_i$ is the moment of inertia of particle $i$
calculated for a spherocylinder with a homogeneous mass density.

The equations of motion are combined with the FIRE
algorithm~\cite{PhysRevLett.97.170201} to minimize potential
energy. In this algorithm, velocities are rescaled after each
time-step to guide the descent in the potential energy landscape.

We have set $k_n=1$ and $c_n=0$. All particles have the same mass
$m=1$ and aspect ratio $\alpha=\ell/d$. Half of the particles have
$d=1$, the other half have $d=1.4$. System sizes are chosen such that
the linear dimension of the simulation box is at least three times the
length of the simulated SC. In terms of particle number this means,
$N=3072\ldots 6144$. The unit of energy is thus $k_nd^2$, times are
expressed in units of the elastic collision time $\sqrt{m/k_n}$.

\subsection{Numerical implementation}

We integrate the equations of motion on a GPU using a velocity Verlet
algorithm for the translational degrees of freedom, and a
Richardson-like iteration for the rotational degrees of freedom (see
appendix), which are represented as quaternions. Normalization of the
quaternions is ensured by rescaling at each time step.

The dynamics is stopped, when the potential energy does not change
appreciably ($\Delta E< 10^{-6}$) and the kinetic energy is below a
threshold ($E_{kin}/N< 10^{-12}$). Most of the time the residual
kinetic energy is many orders of magnitude smaller than this
threshold.

In exceptional cases it may happen that the shortest distance between
two nearly parallel spherocylinders jumps discontinuously from one end
to the other. This may lead to oscillations which make it impossible
to drain the kinetic energy from the system. For nearly parallel SC we
thus include a modification of the distance calculation as follow. The
shortest distance betweeen parallel SCs are taken at the center of the
overlap region. For nearly parallel SCs (angle
$\theta<\theta_{\rm thres}$) this location is linearly interpolated to
the actual shortest distance. As a result no discontinuity arises, and
we find immediate relaxation of kinetic energy of previously
oscillating systems.

\section{Results}

The pressure tensor is calculated from the virial expression
\begin{equation}\label{eq:p_tensor}
  P_{\alpha\beta} = \frac{1}{V}\sum_{k<l} f^\alpha_{kl}R_{kl}^\beta
\end{equation}
where V is the volume of the system, $f^\alpha_{kl}$ is
$\alpha$-component of the force applied on particle k by particle l
and $R^\beta_{kl}$ is the $\beta$-component of the distance between
the particles' center of mass. Using the center-of mass coordinates in
the pressure tensor is not immediately obvious. Such a definition
ensures that the $P_{\alpha\beta}$ is the force in $\alpha$ direction
experienced by a hypothetical wall (with normal along $\beta$
direction) that itself consists of spherocylindrical particles. In a
previous publication on short SC~\cite{nath19:_rheol} we have used the
${r}_{kl}^\beta$ instead. For short particles the difference between
both definitions is very small.

For the isotropic pressure
$p={\rm tr} P/3=
\frac{1}{3V}\sum_{k<l}\mathbf{f}_{kl}\cdot\mathbf{R}_{kl}$ the
difference between both definitions vanishes for side-contacts, as
$\mathbf{f}_{kl}\perp \mathbf{\hat n}_k, \mathbf{\hat n}_l$ and thus
\begin{eqnarray}\label{eq:dotproduct}
 \mathbf{f}_{kl}\cdot\mathbf{r}_{kl}=\mathbf{f}_{kl}\cdot\mathbf{R}_{kl}\qquad\text{(side
  contacts)}\,,
\end{eqnarray}
see Eq.~(\ref{eq:rij}). Below we will find that side-contacts are
dominant whenever SCs are long enough.

\begin{figure}[b]
  \includegraphics[width=0.4\textwidth]{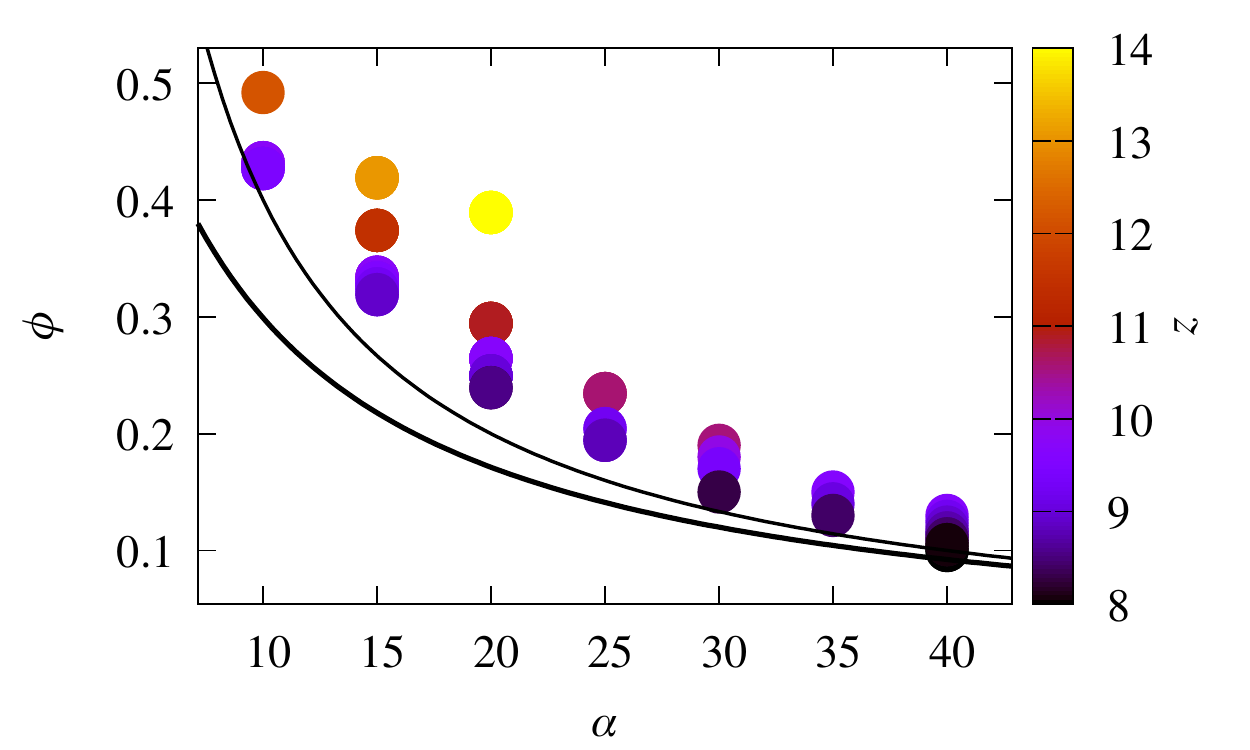}
  \caption{State-points in the plane spanned by volume-fraction $\phi$
    and aspect-ratio $\alpha$. Color code given by 
    number of contacts per particle $z$ of the packing. Thin line is
    $\phi = 4/\alpha$, thick line
    $\phi=\left(1+\frac{\alpha^2/4}{\alpha+2/3}\right)^{-1}$ following
    from Eq.~(\ref{eq:z_vexcl}).}
  \label{fig:phasediagram}
\end{figure}

\subsection{Jammed configurations}

With the procedure just described a set of jammed packings is
generated, \xx{starting from a spatially random initial
  distribution. Effects from ordering can be neglected as particles
  are orientationally constrained and don't move much during
  minimization.} The aspect ratio is varied from $\alpha=10\ldots 40$
and the associated volume fractions are chosen to approach the jamming
threshold. Several packings at the same state point serve to estimate
fluctuations.

Fig.~\ref{fig:phasediagram} shows the state-points of
the generated packings in the $\alpha-\phi$ plane and highlights the
broad range of $\phi$-values necessary to cover the jamming transition
for all aspect ratios. In previous work, Philipse~\cite{phi96} argued
that, asymptotically ($\alpha\to\infty$), the jamming density is
inversly proportional to the aspect ratio, $\phi_J=c/\alpha$. The
proportionality factor $c$ is given by $c\approx z/2$, where $z$ is
the connectivity, the average number of contacts per particle. Thus,
longer SCs do not make more contacts than shorter SCs, but pack at
lower density.

In the general case, the Philipse argument relates the number of
contacts $z$ of a SC to the number density $\rho=\phi/V_{\rm sc}$ and
the orientationally averaged excluded volume~\cite{phi96,onsager49}
\begin{eqnarray}
  V_{\rm excl} &=& (\pi/2)\ell^2d+2\pi d^2\ell+(4/3)\pi d^3\,
\end{eqnarray}
as
\begin{eqnarray}\label{eq:z_vexcl}
z = \rho V_{\rm excl} = \phi \frac{V_{\rm excl}}{V_{\rm sc}}\,.
\end{eqnarray}
This reduces to the above relation $z=2\phi\alpha$ in the limit of
long SCs, where the end caps are irrelevant. Both, the general
expression Eq.~(\ref{eq:z_vexcl}) and the asymptotic version are drawn
in Fig.~\ref{fig:phasediagram} for $z=8$.

At jamming the number of contacts are constrained by mechanical
equilibrium. Maxwell counting~\cite{calladine78} for SC particles with
three translational and two rotational degrees of freedom (no rotation
around the long axis) gives $z_J=10$ contacts per particle that are
minimally necessary to ensure mechanical equilibrium.

\begin{figure}[t]
  \includegraphics[width=0.33\textwidth]{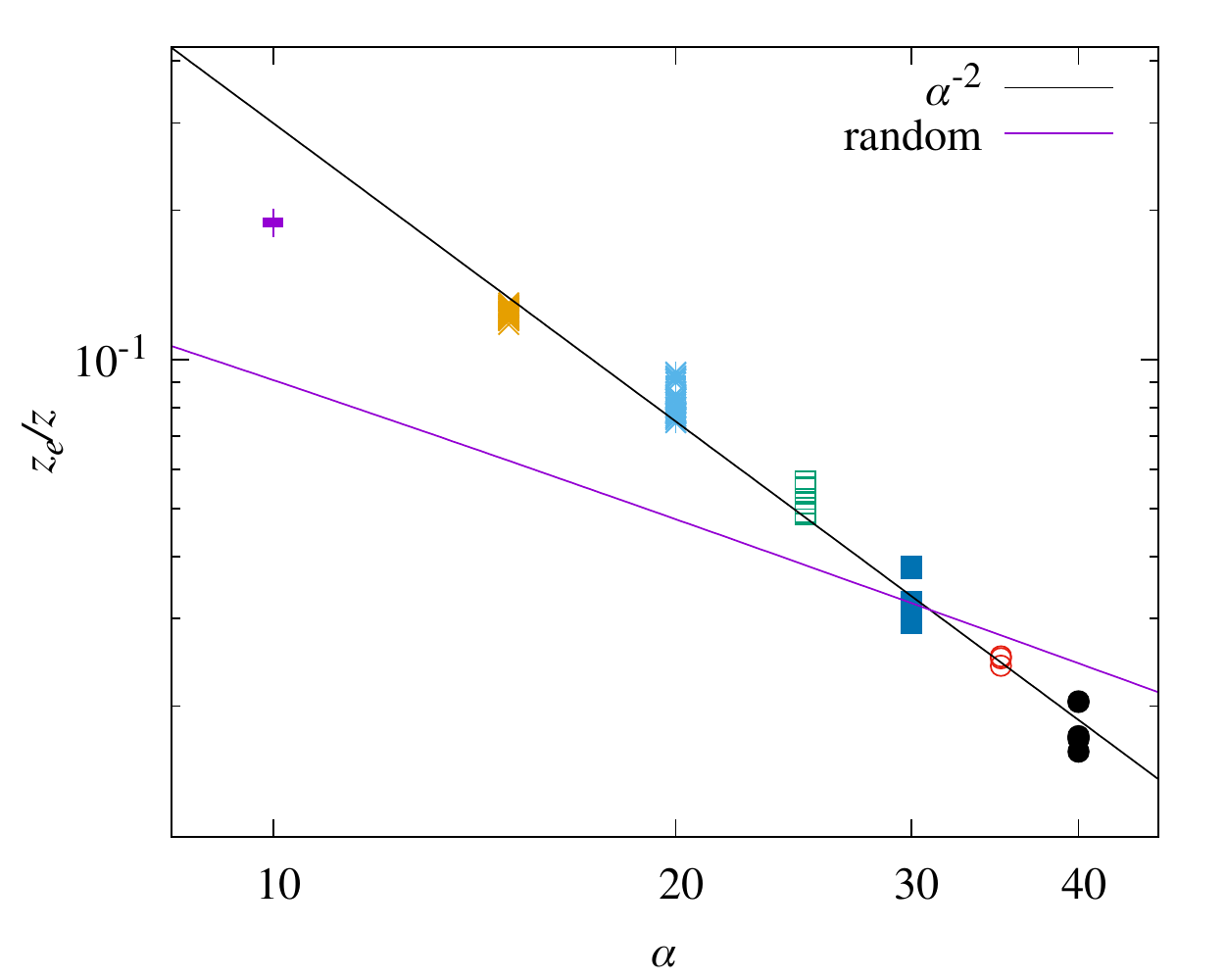}
  \caption{Ratio of end-to-total contacts $z_e/z$ vs. aspect ratio
    $\alpha$. random: ratio of side-to-total area.}
  \label{fig:zside.p}
\end{figure}

\begin{figure*}[t]
  \includegraphics[width=0.32\textwidth]{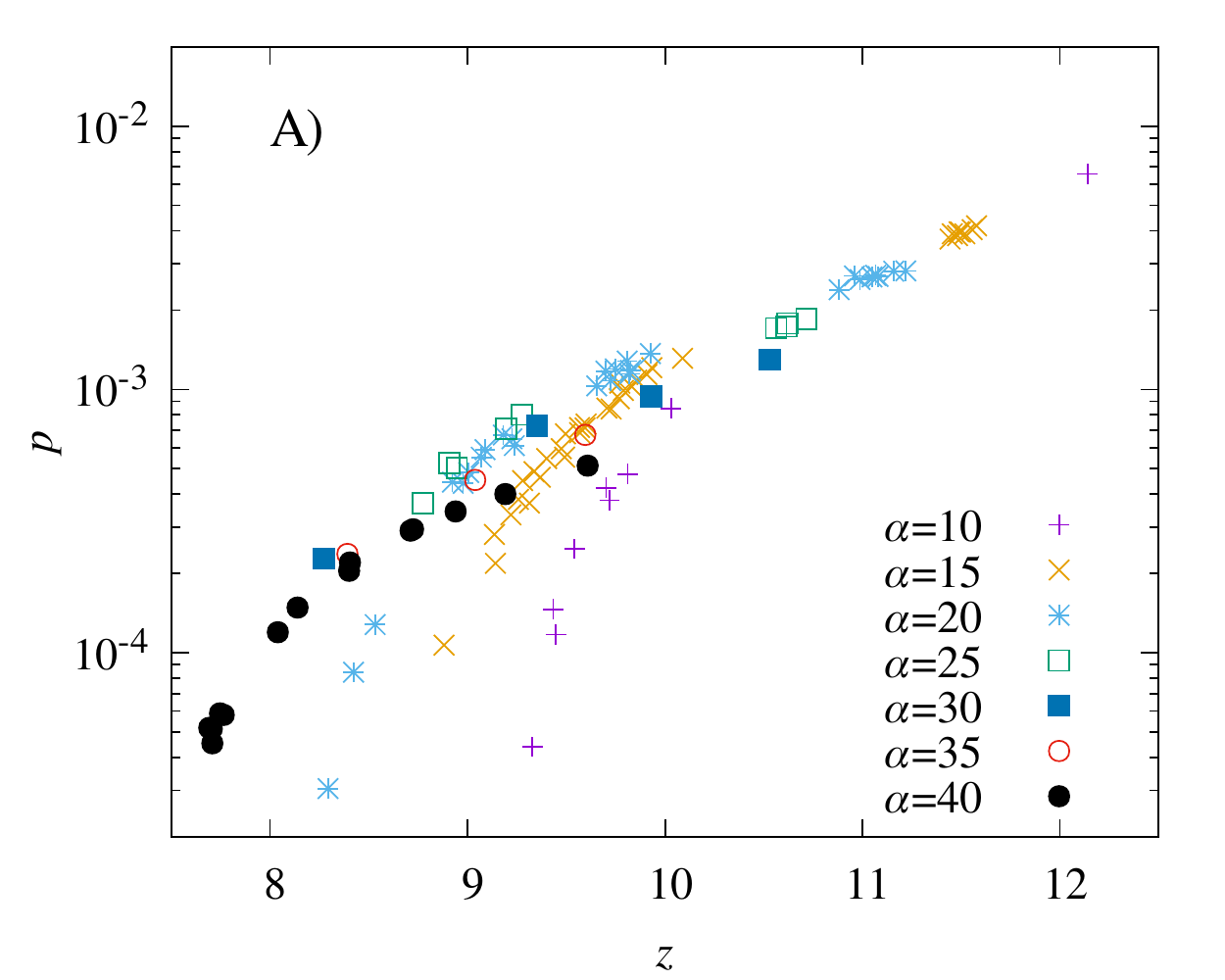}\hfill
  \includegraphics[width=0.32\textwidth]{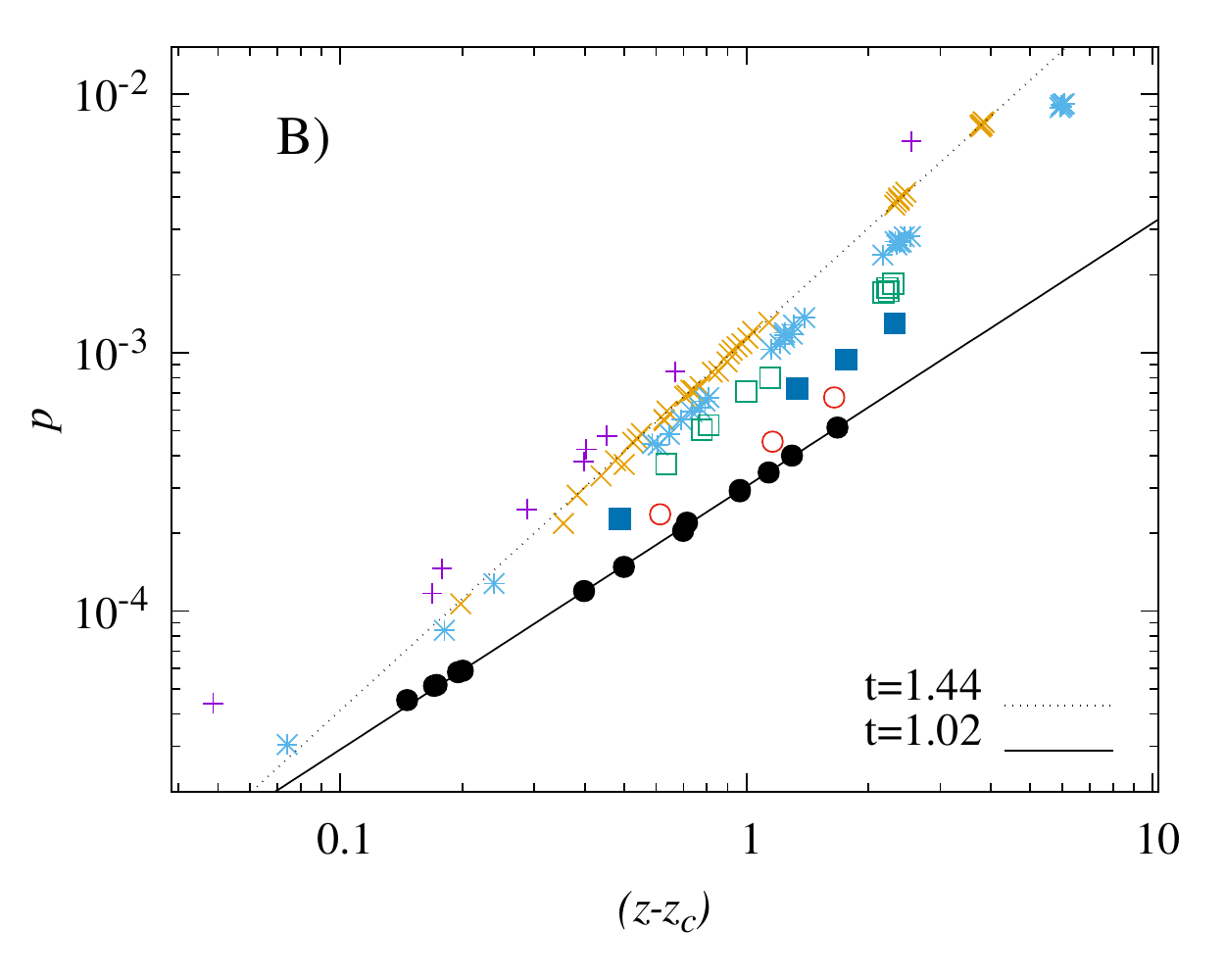}\hfill
  \includegraphics[width=0.32\textwidth]{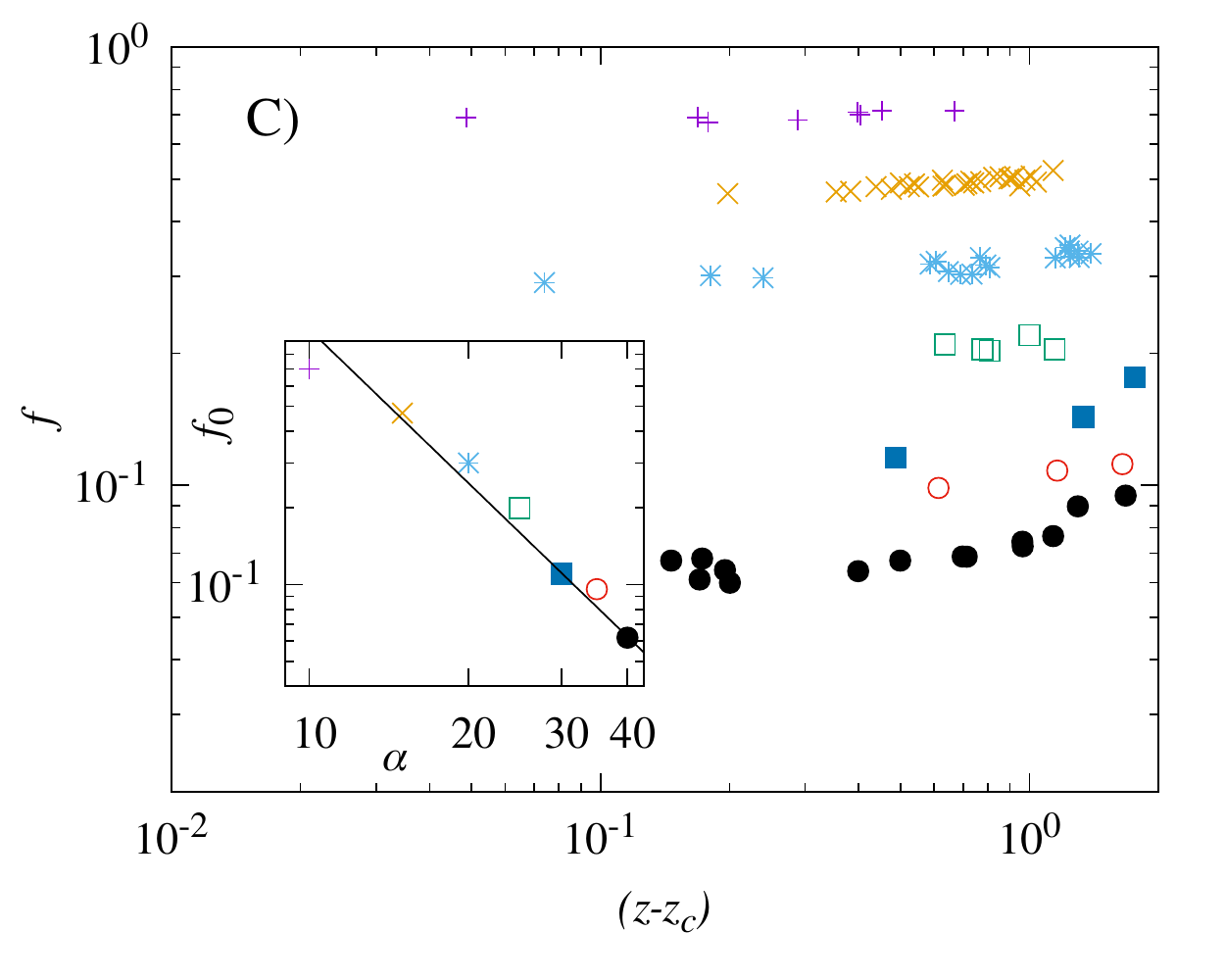}
  \caption{A) Pressure $p$ vs. connectivity $z$ for different aspect
    ratios $\alpha=10\ldots40$ \xx{(rattlers are not accounted
      for)}. B) Pressure $p$ vs. reduced connectivity $z-z_c$ as
    defined in Eq.~(\ref{eq:zc}). The lines are fits of the form
    $p\sim(z-z_c)^t$ to the $\alpha=15$ and $40$ data,
    respectively. C) Fraction of particles with end contacts, $f$,
    vs. reduced connectivity $z-z_c$ for different aspect ratios
    $\alpha=10\ldots40$. Inset: Limiting value, $f_0=f(z\to z_0)$
    vs. aspect ratio $\alpha$. Line is $f_0\sim\alpha^{-2}$. }
  \label{fig:z.p}
\end{figure*}

In fact, the axial translational degree of freedom needs special
consideration in long SCs. It can only be constrained by contacts at
the SC ends, but not by contacts at the sides. As the total number of
contacts does not increase with $\ell$, the relative importance of the
end contacts is expected to vanish. Fig.~\ref{fig:zside.p} shows how
the fraction of end contacts $z_e/z$ decreases with aspect ratio
(roughly as $\alpha^{-2}$) in our different packings. Interestingly,
if contacts were distributed randomly over the entire surface (or
volume) of a SC, the number of end contacts would show a different
behavior, $z_e\sim \alpha^{-1}$, which is also indicated in the
figure.
One can thus conclude that asymptotically, the axial translation mode
is not constrained, thus Maxwell counting gives $z_J=8$, which is the
value used in Fig.~\ref{fig:phasediagram}.

To assess the usefulness of this limit to our packings, we calculate
the pressure and relate it to the corresponding connectivity, see
Fig.~\ref{fig:z.p}a. At jamming, $z=z_J$, the pressure should
vanish. Obviously, the data is a mess and neither prefers $z_J=8$ nor
$z_J=10$, but rather an aspect-ratio dependent $z_c(\alpha)$. \xx{In
  the following we use $z_c$ (and not $z_J$) whenever the
  $\alpha$-dependent threshold $z_c(\alpha)$ is meant.}

\xx{For the longest SCs a value $z<8$, i.e. even below Maxwell
  counting, is observed. This also happens for spherical particles and
  is a sign of the occurence of rattling
  particles~\cite{heussingerPRL2009}. Thus, a more refined analysis is
  warranted.} In the general case, a certain number of particles,
$N_{\rm ec}$ has end contacts (on both ends)~\footnote{Spherocylinders
  with contacts at only one end do hardly ever occur in equilibrated
  states. During minimization the SC can easily (without resistance)
  slide in the opposite direction to remove these
  contacts}. Accounting in addition for rattling particles $N_{\rm
  r}$, i.e. those that do not have any contact (roughly $1\ldots5\%$),
Maxwell counting gives
\begin{eqnarray}\label{eq:zc}
z_c = 8+\frac{2N_{\rm ec}}{N-N_{\rm r}}= 8+2f\,,
\end{eqnarray}
with $f$ the fraction of SCs with end contacts. This fraction is
plotted in Fig.~\ref{fig:z.p}c. As expected, longer SCs have lower
$f$. Apparently, for each aspect ratio $\alpha$ there is a finite
limit $f_0(\alpha)$ when approaching the jamming transition,
$f_0\equiv f(z\to z_c)$. This is plotted in the inset of the
figure. The solid line indicates a dependence
$f_0\propto \alpha^{-2}$, similar to the fraction of end-contacts in
Fig.~\ref{fig:zside.p}.


In Fig.~\ref{fig:z.p}b pressure is plotted again, now against
$\delta z\equiv z-z_c(f)$. The data appears well ordered and even follows
power-laws. The longest SCs $\alpha=40$ seem to suggest direct
proportionality between pressure and contacts, giving $p\sim
(z-z_c)$. However, it seems that the exponent of the power-law is
continuously shifting with aspect ratio. This is rather
unusual. Instead, a cross-over from one power-law to another is to be
expected. \xx{This behavior is because pressure is a combination of
  different factors the crucial one being (see
  Eq.~(\ref{eq:p_tensor})) the normalized sum over contacts
  $1/V\sum_{\rm contacts}(\ldots)$, which can be written as
  $(\phi z/\ell)\langle\ldots\rangle_c$. The observable in brackets is
  related to the overlaps in the contacts. Thus, pressure variations
  reflect different contributions from changing overlaps, but also
  from $z$ and $\phi$, which change appreciably in our ensemble of
  configurations.}

\begin{figure}[ht]
  \includegraphics[width=0.4\textwidth]{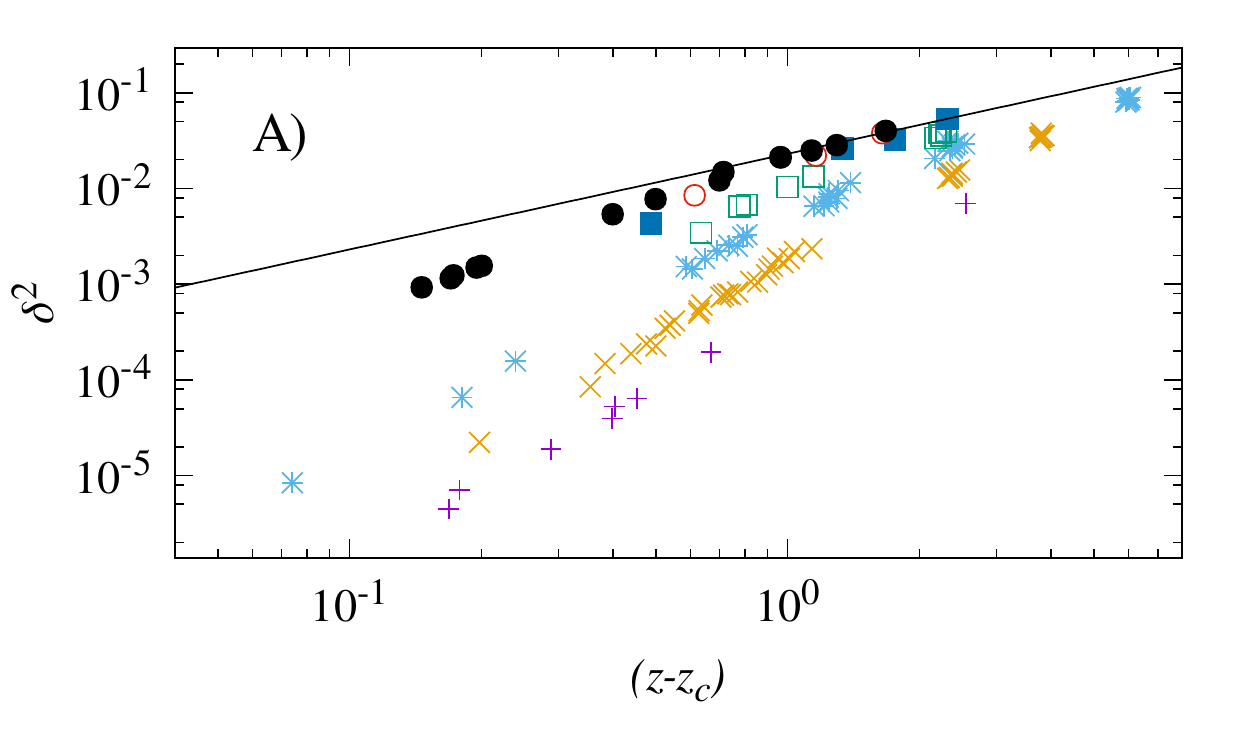}
  \includegraphics[width=0.4\textwidth]{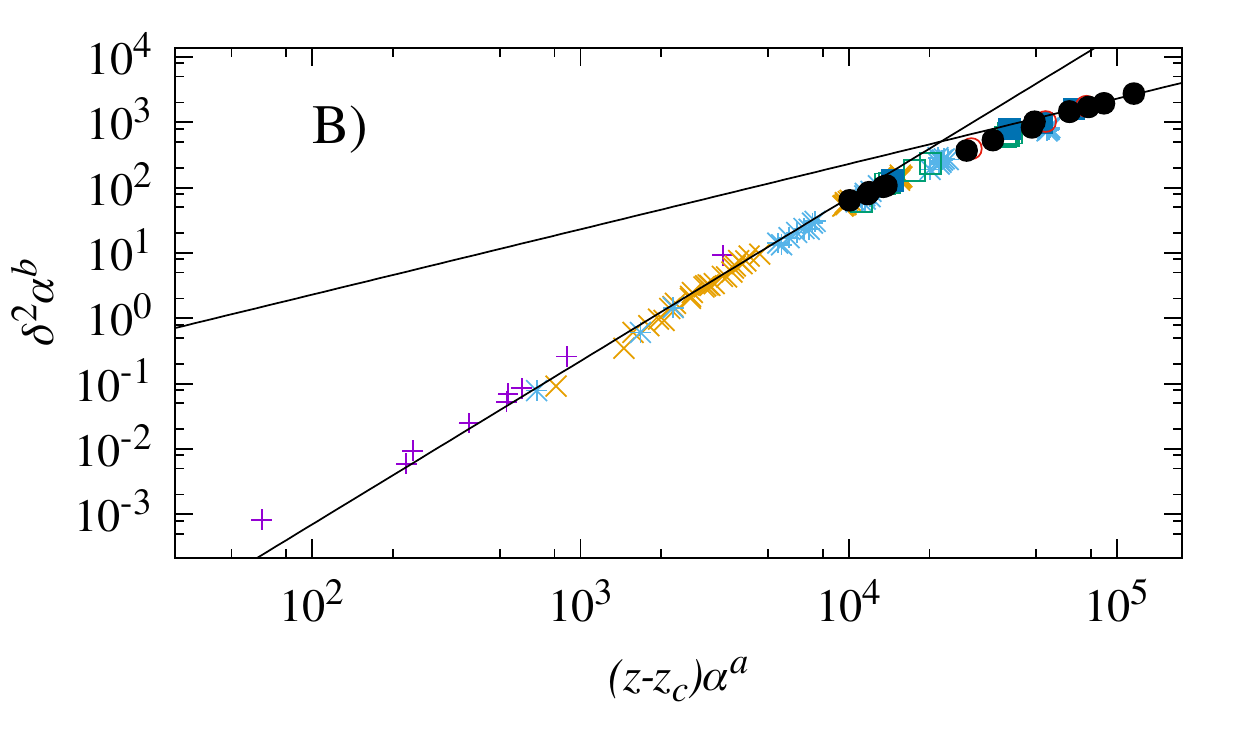}
  \caption{A) Mean square overlap $\langle\delta^2\rangle$ vs. reduced
    connectivity $(z-z_c)$ for various $\alpha$; color code as in
    Fig.~\ref{fig:z.p}. Line is $\delta z^1$. B) Axes rescaled by powers
    of $\alpha$; exponents are determined as $a=b=3$. Lines are
    $\delta z^1$ and $\delta z^{2.5}$.}
  \label{fig:epot.p.scaled}
\end{figure}

%
%
%
\xx{In order to access the properties of the overlaps themselves, one may
  define a rescaled pressure $\hat p = p\ell/z\phi$.
Alternatively, one can study the potential energy (per
contact) $E/N_c=\frac{k}{2}\langle\delta^2\rangle$. Its variations
also directly reflects the overlap via the second moment
$\langle\delta^2\rangle$ of the distribution of overlap values.} In Fig.~\ref{fig:epot.p.scaled}A the
mean-squared overlap as derived from the potential energy is plotted.
The line $\propto (z-z_c)$ points to a regime where
$\langle\delta^2\rangle\propto \delta z$ and independent of SC
length. This regime sets in above a connectivity scale
$\delta z^\star\sim \alpha^{-3}$, as the scaling analysis in panel B
shows. There, we plot the scaling ansatz
$\langle\delta^2\rangle \sim \alpha^{-b}F_\delta(\delta z\alpha^a)$,
with a scaling function $F_\delta$ and exponents $a=b=3$.  \xx{As
  mentioned above, also a rescaled pressure may be used to study the
  overlaps. In order to access the first moment $\langle\delta\rangle$
  a special pressure $p_s$ needs to be defined that is calculated from
  the side contacts only
 \begin{eqnarray}\label{eq:ps_delta}
   p_s = \frac{1}{3V}\sum_{\text{side c.}}\mathbf{R}_c\cdot\mathbf{f}_c\sim \frac{z_s\phi}{\ell}\langle\delta\rangle\,,
 \end{eqnarray}
 where we have used Eq.~(\ref{eq:dotproduct}), which is only
 valid for side contacts.
We have checked that corresponding scaling
properties emerge from the rescaled pressure $\hat p_s\equiv p_s\ell/z_s\phi\sim
\langle\delta\rangle\sim \sqrt{\langle\delta^2\rangle}$ as from the
potential energy.} The effective
\xx{exponents $t$} seen in Fig.~\ref{fig:z.p}B can then be
understood from the dependence of the overlap $\langle \delta\rangle$
together with the variation of the prefactor $z=z_c+\delta z$.

\subsection{Linear response to deformation}

\subsubsection{Bulk modulus}

To probe the response to compressive deformations, a quasistatic
compression protocol is followed. Starting with a minimized packing,
the volume of the simulation box is changed, followed by another
minimization. This is repeated several times to be able to record a
pressure-strain relation. The strain increment $\Delta\gamma$ is
defined from the change of the volume as $\Delta\gamma =-\Delta V/V$,
or in other words, $dV/d\gamma = -V$. The modulus $K$ is defined from
the slope of the pressure-strain relation, $p(\gamma)= K\gamma$, which
is identical to the usual definition of the inverse compressibility
$1/\kappa = -Vdp/dV = -V(dp/d\gamma) (d\gamma/dV) = K$. The strain
values are chosen small enough such that $p(\gamma)$ is a linear
function. It turns out that using $\Delta\gamma= O(10^{-5})$ is small
enough to obtain $5\ldots 10$ points over which the function is indeed
linear.

\begin{figure}[ht]
  \includegraphics[width=0.4\textwidth,clip=true]{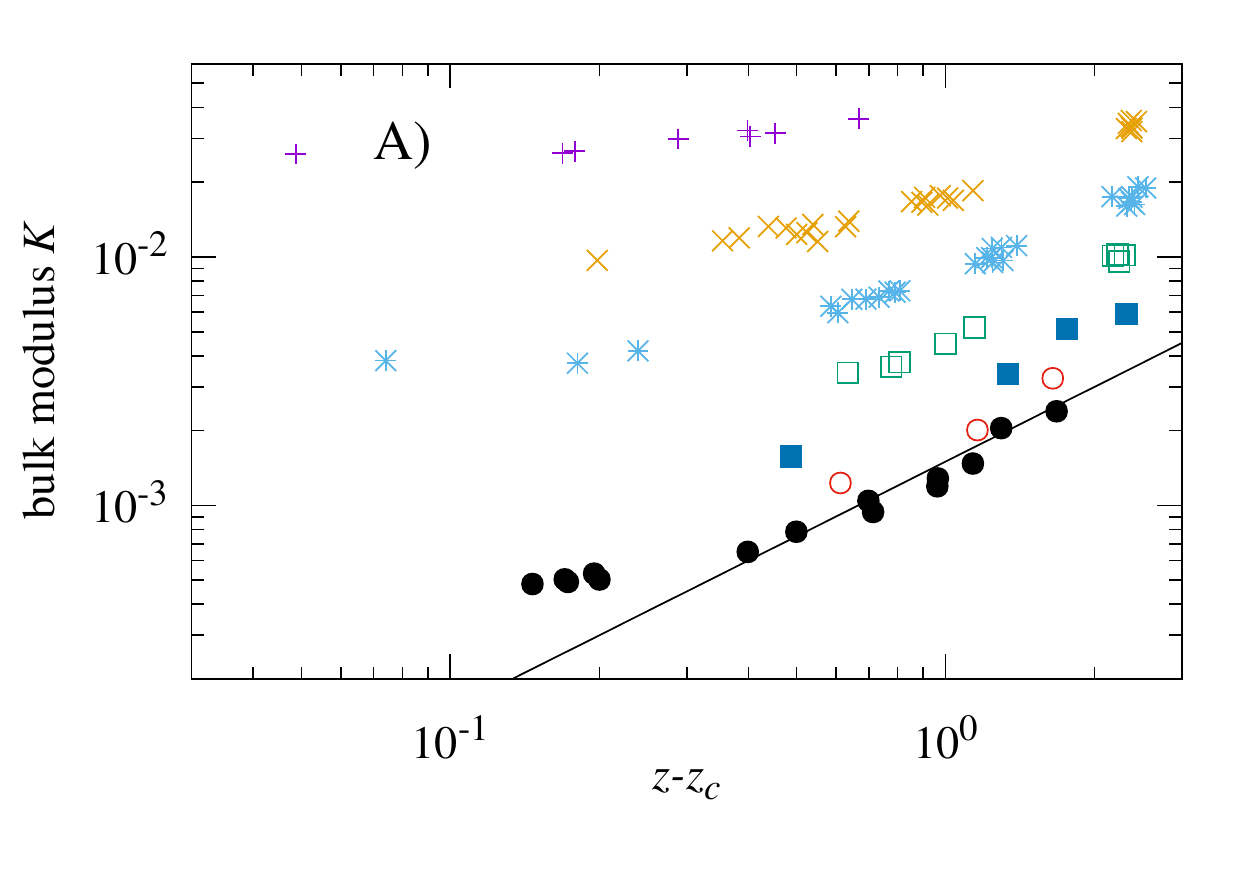}\hfill
  \includegraphics[width=0.4\textwidth, clip=true]{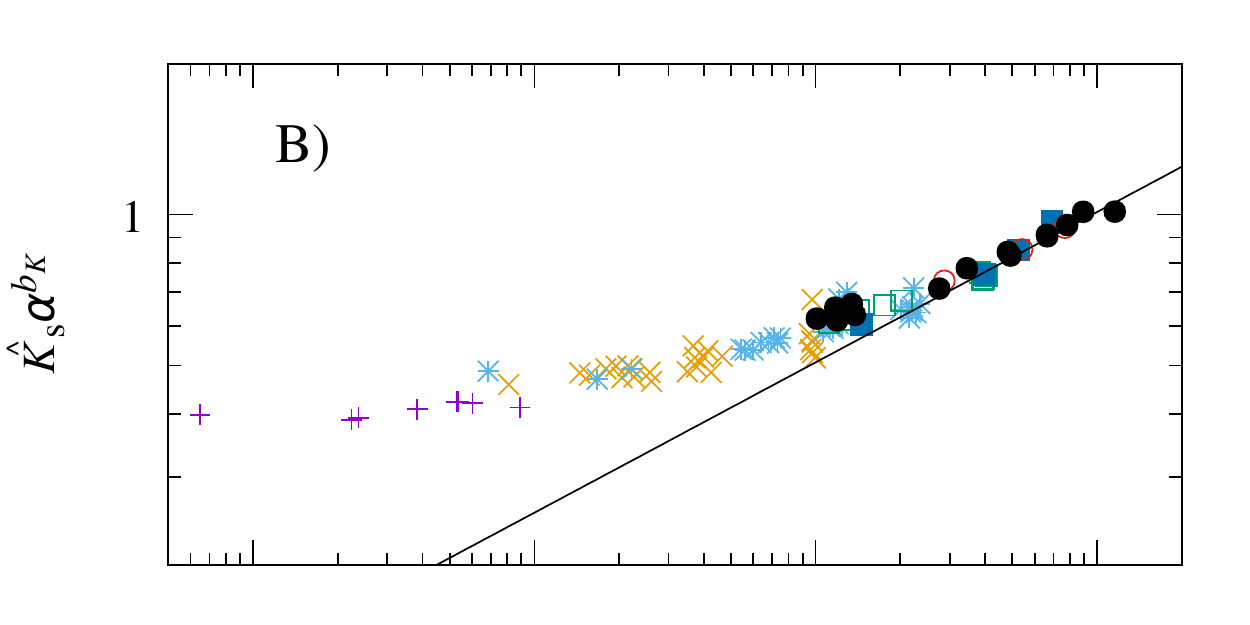}
  \includegraphics[width=0.4\textwidth, clip=true]{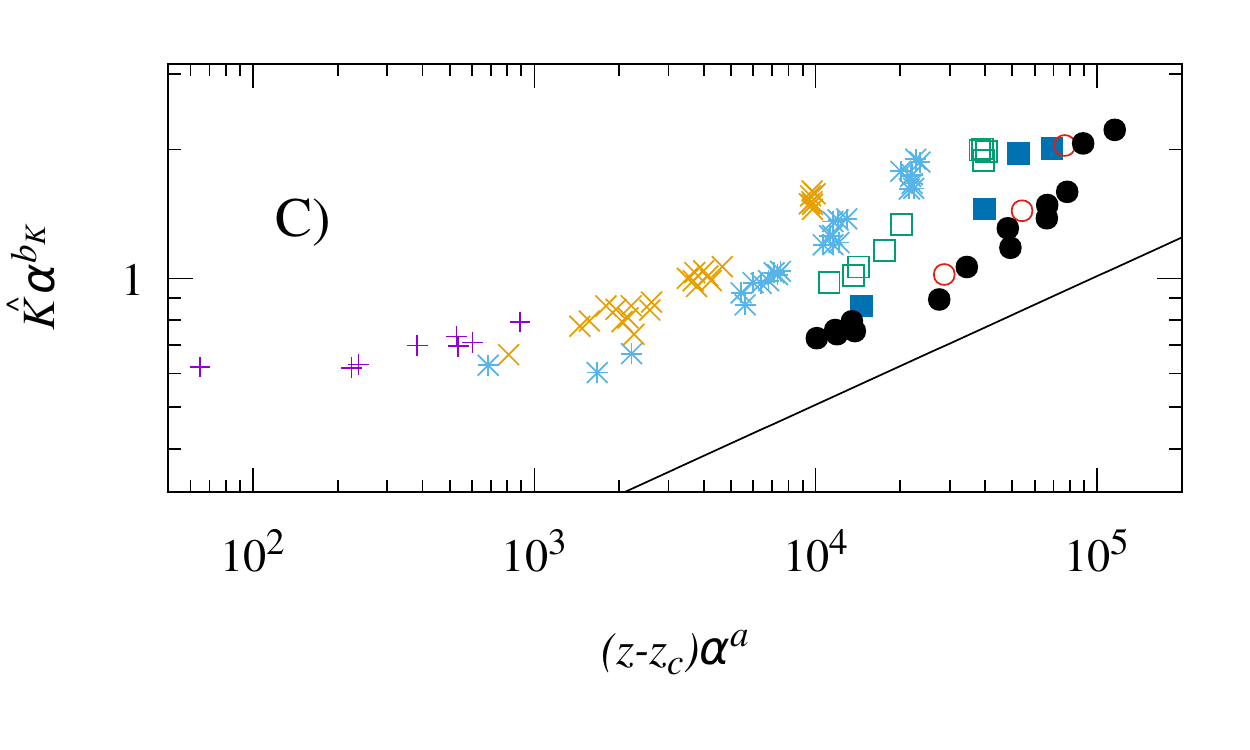}
  \caption{A) Linear bulk modulus $K$ vs. reduced connectivity
    $z-z_c$. Line is $K=0.0015(z-z_c)$. \xx{Color code as in
      Fig.~\ref{fig:modulus.z}. Aspect ratios $\alpha=10\ldots 40$
      from top to bottom.} B) Reduced bulk modulus
    $\hat K_s=K_s\ell/\phi z_s$ vs.  $z-z_c$ and rescaled by powers of
    $\alpha$, $a=3$, $b_K=0.9$. $K_s$ is the contribution to the bulk
    modulus from the side contacts. \xx{Line has slope of $0.3$, see
      text for details.} \xx{C) Reduced (full) bulk modulus
      $\hat K=K\ell/\phi z$ vs.  $z-z_c$ and rescaled by powers of
      $\alpha$. The same exponents are used as in panel B), $a=3$,
      $b_K=0.9$, but no collapse is achieved. Line as in panel B).}}
  \label{fig:bulk.z}
\end{figure}

Fig.~\ref{fig:bulk.z}A plots the bulk modulus $K$ vs. the reduced
connectivity. For large $z-z_c$ the modulus increases linearly in
$z-z_c$. For small $z-z_c$ the modulus reaches an $\alpha$-dependent
constant value $K_0$. This plateau decreases strongly with particle length,
roughly as $K_0\sim\alpha^{-3}$. At least part of this $\alpha$-dependence
stems from the normalization of pressure with volume,
Eq.~(\ref{eq:p_tensor}). As has been derived in
Eq.~(\ref{eq:ps_delta}), the $1/V\sum_{\rm contacts}$ turns into
$\phi z/V_{\rm sc}\to z^2/\alpha^2$. However, the dependence of the
bulk modulus is stronger than this factor $\alpha^{-2}$.

In analogy to Eq.~(\ref{eq:ps_delta}) a reduced bulk modulus of the
side contacts is defined as $\hat K_s=K_s\ell/\phi z_s$, where $K_s$
is the contribution to the bulk modulus from the side contacts,
only. From the definition of the bulk modulus it is clear that it
directly represents the overlaps and their changes under compressive
deformations. This reduced modulus is plotted in
Fig.~\ref{fig:bulk.z}B, where also the axes are scaled by powers of
$\alpha$. Collapse is achieved with a scaling function $F_K$ and $\hat
K_s=\alpha^{-b_K} F_K(\delta z\alpha^a)$, where $a=3$ as for
Fig.~\ref{fig:epot.p.scaled} and $b_K=0.9$. The latter value is the
missing factor that yields $K_0\sim \alpha^{-2.9}$ as also observed in
the full bulk modulus. Notably, the full bulk modulus cannot be scaled
in this way \xx{(see Fig.~\ref{fig:bulk.z}C)}.
Beyond the plateau the data suggest a dependence $\hat K_s \sim \delta
z^{0.3}$. This would also imply that asymptotically the
$\alpha$-dependence drops out, $\hat K_s=\alpha^{-0.9} F_K(\delta
z\alpha^3)\to \alpha^{-0.9}(\delta z\alpha^3)^{0.3}=\delta z^{0.3}$.

\subsubsection{Shear modulus}

Analogously to the bulk deformations, steps of small shear strains
$\delta\gamma=10^{-5}$ are applied via Lees-Edwards boundary
conditions~\cite{lee_edward}. The shear stress $\sigma=P_{xy}$ from
Eq.~(\ref{eq:p_tensor}) is monitored and its dependence on $\gamma$
fit to a linear function. The stress-strain relation is usually nearly
linear, such that a fit yields the linear elastic shear modulus
$\mu=d\sigma/d\gamma$.

\begin{figure}[ht]
  \includegraphics[width=0.4\textwidth,clip=true]{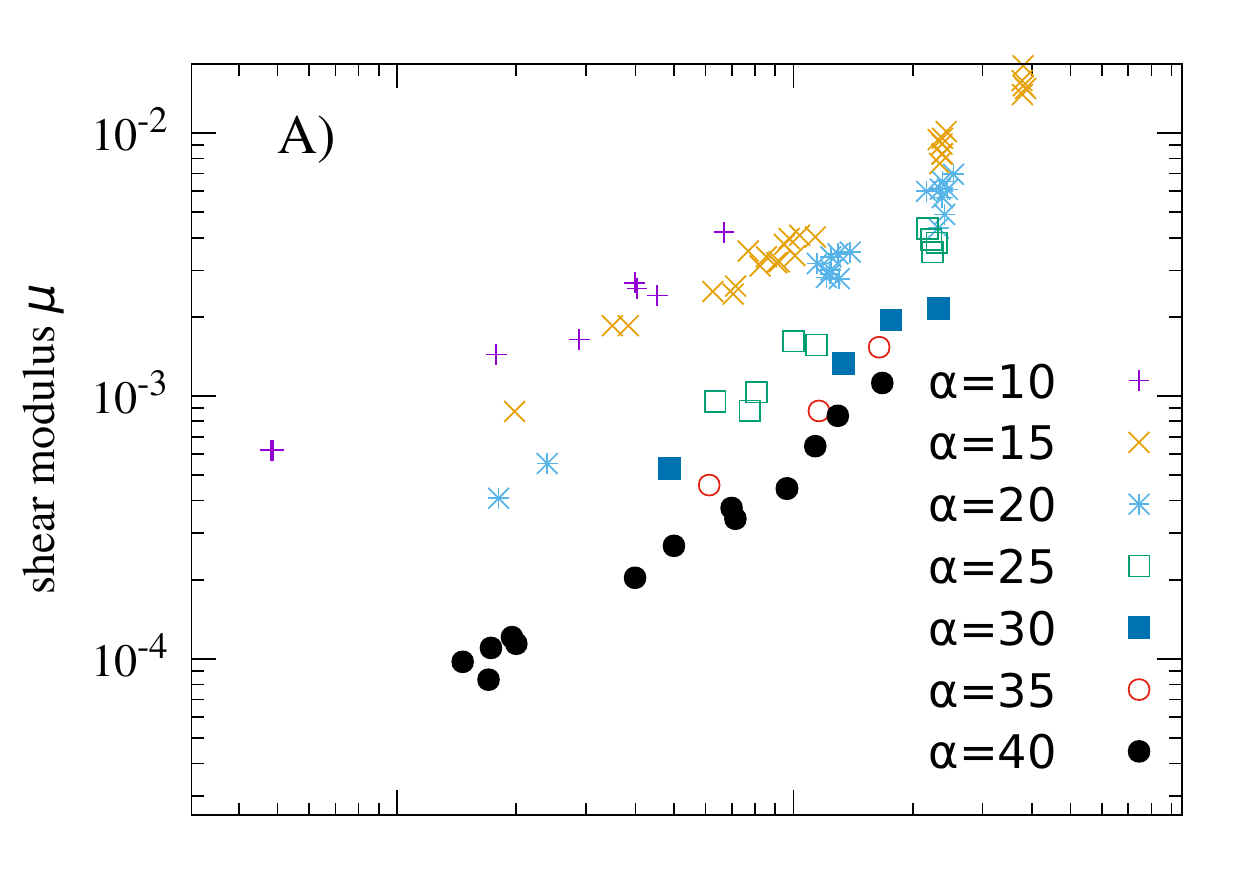}
  \includegraphics[width=0.4\textwidth,clip=true]{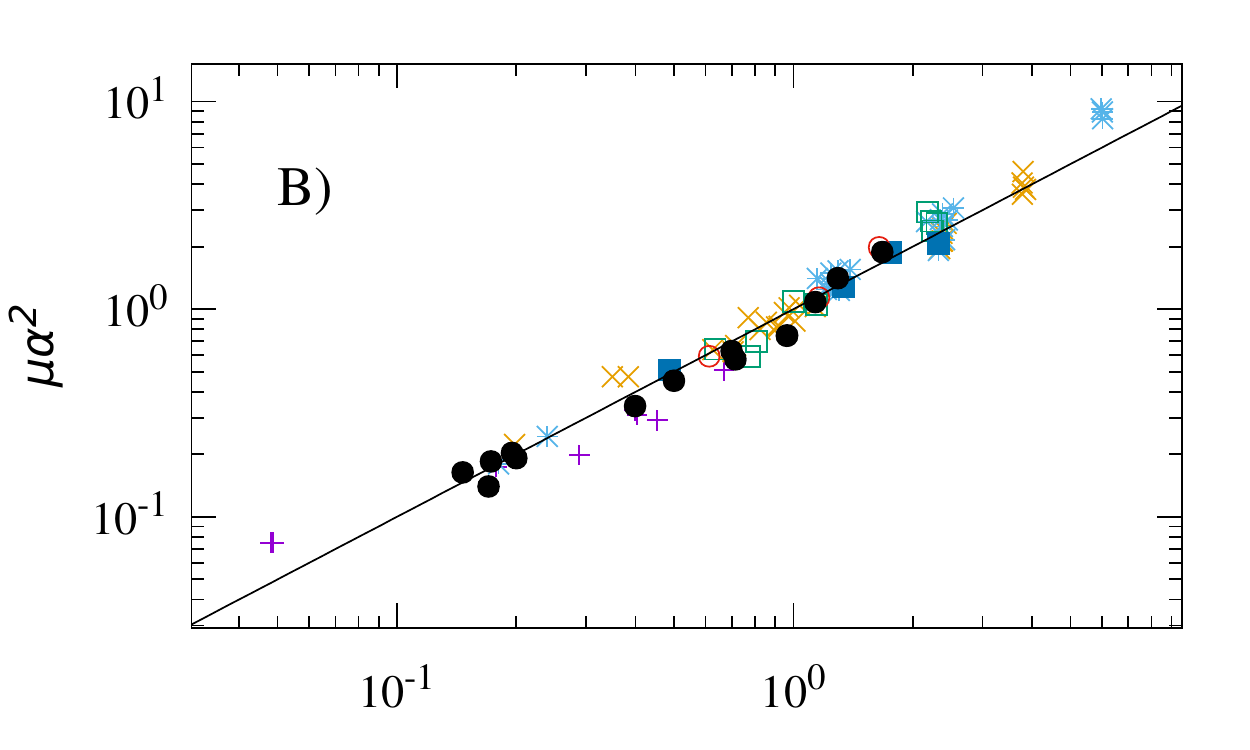}
  \includegraphics[width=0.4\textwidth,clip=true]{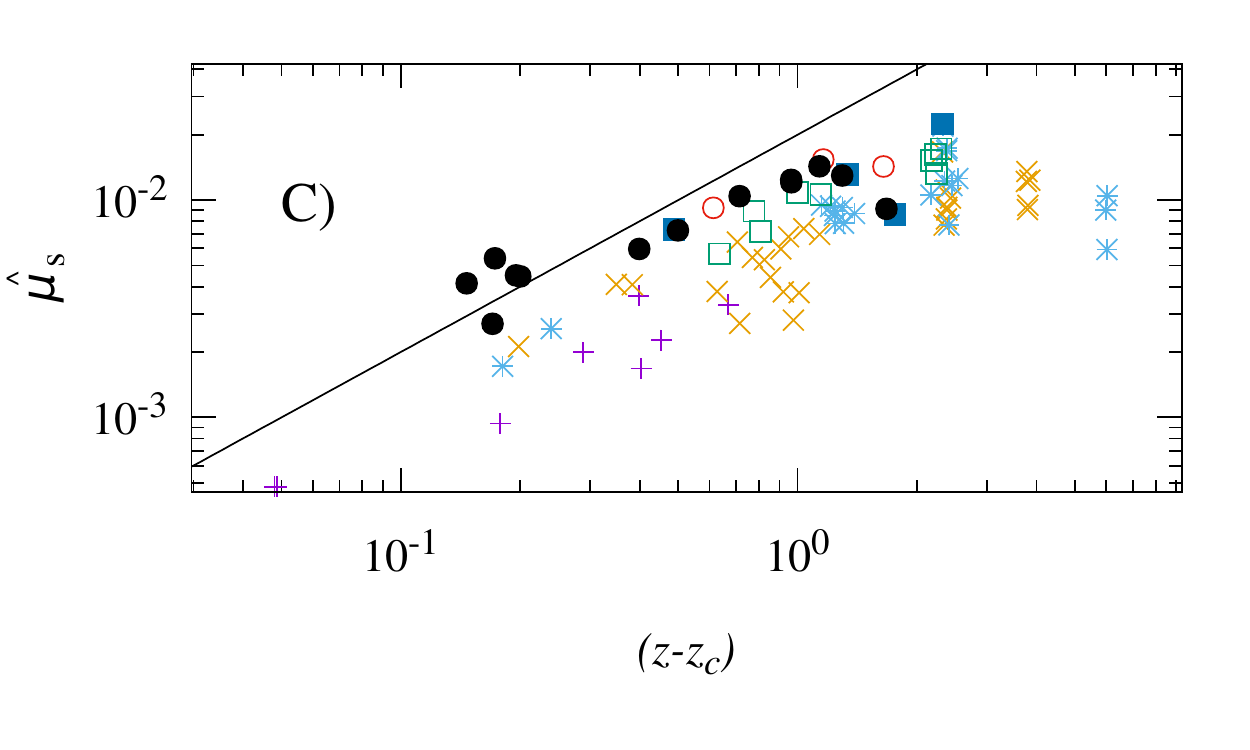}
  \caption{A) Linear shear modulus $\mu$ vs. reduced connectivity
    $z-z_c$. B) Scaled shear
    modulus $\mu\alpha^2$; line is $\mu\alpha^2=(z-z_c)$. C) Reduced
    shear modulus $\hat \mu_s=\mu_s\ell/\phi z_s$ vs.
    $z-z_c$ 
    $\mu_s$ is the contribution to the shear modulus from the side
    contacts. Line has slope of $1$.}
  \label{fig:modulus.z}
\end{figure}

In contrast to the bulk modulus, the shear modulus does not present a
plateau at small $z-z_c$, such that close to jamming $\mu\ll K$. As in
the standard scenario for spherical particles
($\alpha=0$)~\cite{ohern03} we find $\mu/K\sim \delta z$. Panel B) of
the same figure illustrates in more detail the scaling properties of
the shear modulus. By rescaling the y-axis with $\alpha^2$ we can show
that $\mu=\alpha^{-2}(z-z_c)$. For completeness we also display the
reduced shear modulus $\hat\mu_s=\mu_s\ell/\phi z_s$ of the side
contacts $z_s$. Surprisingly, the data is rather noisy, much more so
than the full modulus. For example, the data for $\alpha=15$ (yellow
crosses) show quite some scatter at intermediate $z$, which is much
smaller in the full modulus. Apparently, the splitting into the two
components from end- and side-contacts is quite variable. A large
contribution from the side contacts may be buffered by a small
contribution from the end contacts, and vice-versa.

\section{Discussion}

We have presented simulation results on the properties of packings of
frictionless spherocylindrical particles.

A packing represents a force-balanced, minimal energy state, given
that the spherocylinders (SC) interact via pairwise contact forces at
the point of closest approach.

Spherocylinders have a special shape that gives rise to interesting
properties. First, rotational symmetry around the axis sets the
jamming threshold at $z_J=10$ contacts per SC (two contacts per degree
of freedom). Second, there is an approximate translational symmetry
along the long axis. For contacts at the SC-side interaction forces
are directed perpendicular to the long axis. Therefore translation
along this direction (``sliding'') does not change any of these forces
and no resistance builds up. If this symmetry would be perfect,
jamming would happen at $z_J=8$. Contacts at the SC-end, of course,
break the symmetry as they do feel sliding, and resist such motion.

In our simulations we find that packings of long SC have very few end
contacts, much less than, e.g. a random distribution of contacts on
the surface (or over the volume) of a SC would suggest. The fraction
$f_0$ of particles with ends constrained decreases with SC-aspect
ratio as $f_0\propto \alpha^{-2}$ (see Fig.~\ref{fig:z.p}C). We can
explain this scaling via an extension of Philipse's
argument~\cite{phi96} for the jamming density $\phi_c\sim
z/\alpha$. If SCs are distributed randomly in space, then $f_0$ is
just the probability that two out of $N$ SCs interesect a given test
particle at its two ends.  \xx{If we define $\pi_{1e}$ as the
  probability that the test particle is intersected by \emph{one} SC
  at either end, then $f_0\approx \frac{N(N-1)}{2}\pi_{1e}^2$. The
  probability $\pi_{1e}\approx V_{\rm sc}/V_{\rm box}\ll 1$ as the the
  center of mass of the intersecting SC has to lie within a volume
  $V_{\rm sc}$ close to the end of the test particle.} As a result, we
find $f_0\sim\rho^2V_{\rm sc}^2 \sim\phi_c^2 \sim\alpha^{-2}$ as
observed (to be compared to the probability of single contacts at side
or end $\pi_1=V_{\rm excl}/V_{\rm box}$ and $z=N\pi_1\sim \phi\alpha$,
which is Eq.~(\ref{eq:z_vexcl})). Given, that only a fraction of
particles have their ends constrained, the jamming threshold smoothly
interpolates from $10$ to $8$, when $f_0$ decreases from $1$ to 0. In
fact, one can use constraint counting to derive $z_c(f_0)=8+2f_0$.

These findings suggest a comparison with the system of ellipsoidal
particles~\cite{zeravcic2009EPL,PhysRevLett.102.255501} that has been
mentioned in the introduction. When going from spheres to slightly
aspherical ellipsoids new rotational degrees of freedom are introduced
and constraint counting gives a new jamming threshold $z_J=12$ (in the
general case of three distinct axes). In fact, this threshold is not
immediately reached as the new degrees of freedom are either
zero-energy (quartic) modes or form a separate rotational band. This
only weakly interferes with the jamming threshold. Only at larger
asphericity, when rotational and translational degrees of freedom mix,
do the zero modes vanish and full ellipsoidal jamming is reached.

In the case of spherocylinders it is the lowering of the SC length
(from infinity) that introduces a new degree of freedom, that of
translation along the SC axis. As long as the ends of the SC are not
constrained, this mode is a zero-energy (but not quartic) mode. For
the remaining fraction $f_0$ of SCs with ends constrained this
translational mode is expected to be of finite-energy and corresponds
to the rotational band in the case of ellipsoids. It would be
interesting to compute the density of states to see if these modes
also form a separate band and if so when (at what SC length) mixing is
observed.

The suitable control parameter to measure the distance to jamming then
is $\delta z\equiv z-z_c(f_0)$ with $f_0(\alpha)\sim\alpha^{-2}$. This
defintion allows to easily compare different aspect ratios $\alpha$
that generally jam at wildly different volume fractions
$\phi_J(\alpha)\sim\alpha^{-1}$. Given the control parameter $z-z_c$
we present measurements of pressure $p$ and potential energy (or
mean-squared overlap $\langle \delta^2\rangle$) of the packings, as
well as their response to bulk and shear deformation.

The analysis of the parameter dependence is complicated by the
simultaneous variation of several quantities. As an example consider
Eq.~(\ref{eq:ps_delta}), which is an expression for the part of the
pressure that stems from the side contacts $z_s$,
$p_s\sim \frac{z_s\phi}{\alpha}\langle \delta\rangle$. In the
immediate vicinity to jamming, one can safely set $z_s=z_{sJ}$ and
$\phi=\phi_J$ and neglect their variation with the control parameter
$\delta z$. The pressure then only changes due to changes of the
overlap $\langle\delta\rangle$ with $\delta z$.

As it turns out, we cannot produce packings close enough to jamming to
guarantee this limiting behavior. Equilibration times quickly reach
time-scales that are no longer practical with our simulation
methods. It might be the very sliding motion of end-unconstrained SCs
that spoils equilibration, as this motion cannot be resisted.  This
dilemma is visible when comparing the pressure data in
Fig.~\ref{fig:z.p}B with the averaged overlaps
$\langle\delta^2\rangle$ in Fig.~\ref{fig:epot.p.scaled}. The overlaps
are characterized by a cross-over scale
$\delta z^\star\sim \alpha^{-a}$ with $a\approx 3$. On the other hand,
such a scale is not visible in the pressure data, which rather feature
power-laws with continuously shifting exponents
$t\approx 1\ldots 1.5$. By analyzing $\hat p_s = \alpha p_s/z_s\phi$,
we have verified that the scale is indeed hidden due to variations of
the factors $z$ and $\phi$.

The physical origin of the crossover scale is currently unclear. A
clue to its understanding may lie in the rather large numerical
prefactor $\delta z^\star \approx 2\cdot10^4\alpha^{-a}$, which is
rather unusual. A possible resolution may be to write $\delta z^\star$
in terms of $f_0$ rather than $\alpha$. With the numerical factor
taken from Fig.~\ref{fig:z.p}C (inset), we would obtain
$\delta z^\star \approx 20 f_0^{3/2}$. A negative exponent $-a=-3$
indicates that asymptotically, i.e. for $\alpha\to\infty$, the regime
$\delta z >\delta z^\star$ prevails. For the overlaps, for example,
this means $\langle\delta^2\rangle \sim \delta z$ independent of
$\alpha$. This scaling means, that each of the $N\delta z$ constraints
contributes independently and roughly equally to the potential energy.

A similar scale is observed in the bulk modulus $\hat K_s$ of the side
contacts (see Fig.~\ref{fig:bulk.z}B) which, via its definition
$K_s=dp_s/d\gamma$, represents the change of overlaps with dilational
strain, $\hat K_s\sim d\delta/d\gamma$. In the asymptotic regime the
$\alpha$-dependence drops out and
$\hat K_s=\alpha^{-0.9} F_K(\delta z\alpha^3)\to \alpha^{-0.9}(\delta
z\alpha^3)^{0.3}=\delta z^{0.3}$.

Interestingly, the full bulk modulus $K$, which also includes contacts
from the SC ends, shows different properties. It is linear in
$\delta z$ and cannot be scaled with $\delta z^\star$. It seems
reasonable to suppose that this difference is a consequence of the
special role of the SC length $\ell$ in the end contacts. For end
contacts $\ell$ enters the pressure in a different way than for side
contacts. This is apparent in the definition of pressure
(Eqs.~(\ref{eq:p_tensor})) via the virial contribution of a contact,
$\mathbf{f}_{kl}\cdot\mathbf{R}_{kl}$, where $\mathbf{R}_{kl}$ is the
vector between the center-of-masses of the two contacting SCs $k$ and
$l$, and thus $|\mathbf{R}_{kl}|\sim \ell$. Another way of seeing the
special $\ell$-dependence in the end-contacts is by considering an
affine deformation of the packing. From the properties of an affine
map, the distance between the two ends of a SC (being $\ell$ apart)
should change by $\propto\gamma\ell$. As the SC itself does not change
length, this would naturally induce additional overlaps of exactly
this size. On the other hand, side contacts will only experience
overlaps $\propto\gamma d$, with the diameter $d$ of the SC.

Closer towards jamming the bulk modulus has a plateau which strongly
decreases with SC length. This decrease is stronger than the prefactor
$\frac{z\phi}{\alpha}$ might suggest. Thus, the overlaps themselves,
or rather their change with strain decreases with SC length,
$\delta'\sim \alpha^{-0.9}$.

Finally, we also calculate the shear modulus $\mu$ (see
Fig.~\ref{fig:modulus.z}). In contrast to the bulk modulus the shear
modulus does not have a plateau, but vanishes continuously at jamming,
$\mu\sim \alpha^{-2}(z-z_c)$. For the ratio of both we find $\mu/K\sim
\delta z\to 0$, similar to packings with spherical
particles~\cite{ohern03}. The scaling with $\alpha^{-2}$ is again due
to the prefactor $z\phi/\alpha$. The splitting of the shear modulus in
contributions from side and end contacts does not seem to be
reasonable, as the scatter is unexpectedly high. In fact, the
distinction between side and end contacts is more important for the
pressure as for the shear stress, as Eq.~(\ref{eq:dotproduct}) shows.

A vanishingly small shear modulus has also been observed in packings
of elastic fibers~\cite{PhysRevE.80.016115}. As it turns out, by
introducing frictional interactions between the fibers, the shear
modulus strongly increases~\cite{PhysRevE.80.016115}. As friction
primarily inhibits sliding motion, this supports our understanding
that it is the axial sliding degrees of freedom of the SCs that are
responsible for the increased shear modulus.

A similar effect is observed in bonded networks of fibers, where
permanent bonds take the role of the frictional interactions, and
bond/fiber deformation that of
overlaps~\cite{PhysRevE.93.062502,heussingerPRE2007}. Again, the
length-scale $\ell=\alpha d$ plays a special role. If fibers try to
slide in response to shear, the anchoring points of fiber bonds will
be displaced by an amount $\propto \gamma\ell$. This then is the
amount of strain induced locally either in the bonds or the
fibers. Note, that this strain is much larger than $\propto\gamma d$
with $d$ a microscopic length-scale like the length of a bond or the
distance between bonds.

Future work should analyze these analogies between the different
systems in more detail, in particular paying attention to the
respective role of the ``mesoscopic'' length-scale $\ell$. This length
is on the one hand much larger than the typical microscopic
length-scales as diameter, bond length or mesh-size of the structure,
but also supposedly much smaller than the scale of the entire system,
be it a granular heap, a fiber network or a polymer mesh.

\begin{acknowledgments}
  Financial support by the German Science Foundation (DFG) via the
  Heisenberg program (HE-6322/2) is acknowledged.
\end{acknowledgments}

\begin{appendix}

  \section{Integration of rotational degrees of freedom}

  The integration of the particle orientation utilizes the equation~\cite{rapaport_2004}
  \begin{equation}
    \vec{\dot q} = \frac{1}{2} \vec{W}^{\rm T}(\vec q) \boldsymbol\omega_4\,.
  \end{equation}
  with angular velocities
  $\boldsymbol{\omega}_4 = (\boldsymbol{\omega},0)$ taken in the
  particle frame.

  The updated quaternion $\vec q_{\rm new}$ is obtained in a two step
  process, via
  \begin{eqnarray}
    \vec{q}_1 &=& \vec{q}_{\rm old} + dt\vec{\dot q}|_{\rm old}\\
    \vec{q}_{1/2} &=& \vec{q}_{\rm old} + (dt/2)\vec{\dot q}|_{\rm old}\\
    \vec{q}_{2} &=& \vec{q}_{1/2} + (dt/2)\vec{\dot q}|_{1/2}\\
    \vec{q}_{\rm new} &=& 2\vec{q}_2 -\vec{q}_1\,.
  \end{eqnarray}
  
  The value of $\vec{\dot q}|_{1/2}$ is obtained from
$\boldsymbol\omega_{1/2}$ via the angular momentum $\vec l$ in the lab frame
\begin{equation}
  \vec{l}_{1/2} = \vec{l}_{\rm old} + (dt/2)\vec{t}
\end{equation}
with the torque $\vec{t}$. Transforming to the particle frame
\begin{equation}
\vec I\boldsymbol{\omega}_{\rm 1/2} = \vec{R(\vec{q}_{1/2})}\vec{l}_{1/2}
\end{equation}
where $\vec I$ is the (diagonal) moment of inertia in the particle
frame and $\vec{R}$ is the rotation matrix transforming from lab to
particle frame.

\end{appendix}

%


\end{document}